# Orbital and Physical Parameters of Visual Binary: WDS 17190-3459
## ($A_{2000} = 17^h\ 18^m\ 56^s$ and $\Delta_{2000} = -34^o\ 59'\ 22"$)

***Rukman Nugraha and S. Siregar***
*Astronomy, Faculty of Mathematics and Natural Sciences, Institut Teknologi Bandung*
*Jl. Ganesha No. 10 Bandung 40132*
*e-mail: rukman@students.as.itb.ac.id*

*Abstract*

*Since the Bosscha Observatory was established in 1923 researches on visual binary stars played an important role in astronomical studies in Indonesia. The visual binary of WDS 17190-3459 = MLO 4AB = HD156384 = HIP84709 was extensively observed at our observatory and other observatories. This system has already passed periastron three times since observed in the end of year 1876. The observation data is more than enough to construct an orbit. By using Thiele-Innes method we computed the orbit, and physical parameters are determined by using mass-luminosity relation. The result is presented in the table.*

Table: Orbital and physical parameters of WDS 17190-3459

| Orbital Parameters | | | Physical Parameters | |
|---|---|---|---|---|
| $e = 0.578$ | $i = 132^o.7$ | $a = 1".713$ | $p = 0".134$ | $M_1 = 0.6\ M_o$ |
| $P = 42.3$ years | $\omega = 247^o.5$ | $\mu = 8^o.51$/years | $M_{bol1} = 6.7$ | $M_2 = 0.5\ M_o$ |
| $T = 1974.9$ | $\Omega = 318^o.1$ | | $M_{bol2} = 7.4$ | $q = 0.863$ |

At time being there are several new methods for determining the orbit; for example the method of Gauss done by Söderhjelm (1999) for calculating the orbit of the same stars WDS 17190-3459. Our results are relatively same.

## I. Introduction

Since the Bosscha Observatory was established in 1923 researches on visual binary stars played an important role in astronomical studies in Indonesia. One of these researches is determining its orbital elements and physical parameters.

In this work, we choose WDS 17190-3459 = MLO 4AB = HD156384 = HIP84709 ($\alpha_{2000} = 17^j\ 18^m\ 56^s.36$; $\delta_{2000} = -34^o\ 59'\ 22".5$) due to location of the stars at Shouthern hemisphere, has been extensively observed at Bosscha observatory and other observatories, and have already passed periastron three times since observed in the end of year 1876. It means the observation data is more than enough to construct an orbit.

In order to construct an orbit we use Thiele-Innes method. Siregar (1988) developed a software for calculating orbit and mass of visual binaries based on algorithm for elliptic motions by using this method. In this work, we follow his calculations; so we don't repeat the method used here. Readers interested are suggested to see Siregar (1988). In determining physical parameters we use Stand's Mass-Luminosity relation. This can be done due to visual magnitudes and spectrums are known. The results using Thiele-Innes method and Stand's Mass-Luminosity relation are compared with reference (Söderhjelm, 1999), which use Gauss method.

## II. Orbital Parameters

In Thiele-Innes method, the determination of Kepler's constant C is the first step and most crucial in orbital parameters' determination of visual binary. To do this firstly we corrected the position angles $\theta$ of all of observational data to the equinox 2000.000, as shown in Figure 1.

Figure 1. Observation data of WDS 17190-3459

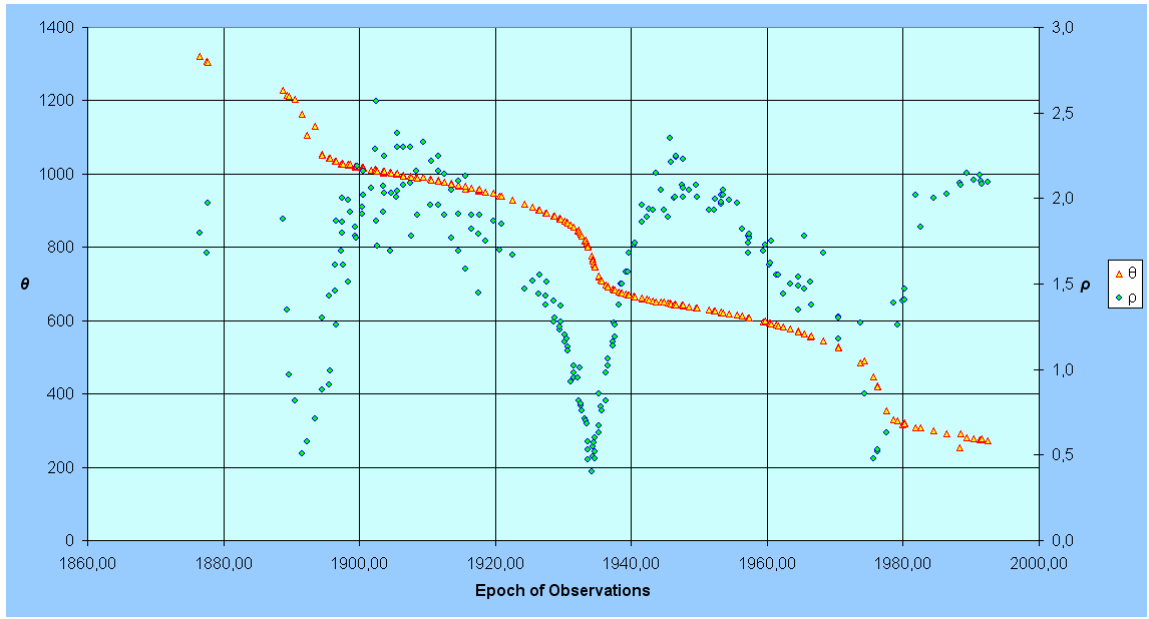

All of these corrected data are interpolated and corrected-by-hand to get $\theta$ and angular separation $\rho$ for fixed times, as shown in Figure 2, and the double of the area constant known as Kepler's II law. We get Kepler's constant - $0.2456 \pm 0.0002$ rad ["]$^2$/years.

Figure 2. Corrected-by-hand of $\theta$ and $\rho$ for fixed times

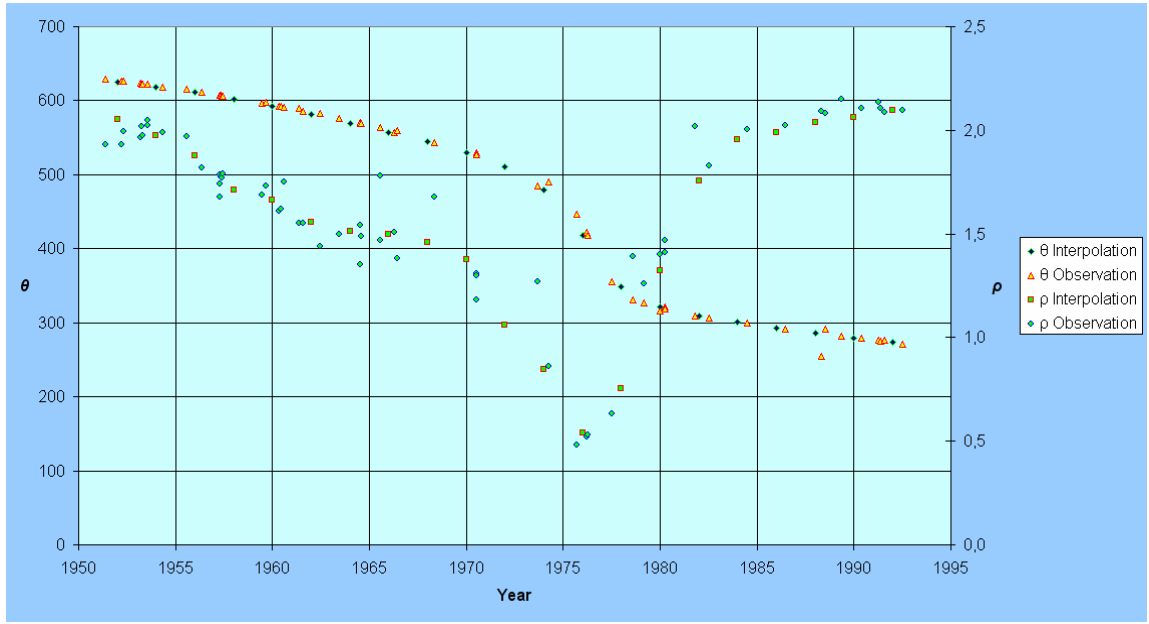

After that we choose three fundamental points that will be used in determining the orbit. We choose 1952, 1972, and 1992, as shown on Table 1. From this we obtain annual mean motion $\mu$ which fulfill $E_{21} + E_{32} = E_{31}$, where $E_i$ is eccentric anomaly for time $i$. Values for $E_1$, $E_2$, and $E_3$ are $170°.465$, $309°.619$ and $517°.802$, respectively. This result yields dinamical elements ($e, P, T$), orientation elements ($i, \omega, \Omega$), and geometric ellips ($a, b$). All of these value are shown in Table 5. We get also Thiele-Innes Constant and its error, as shown in

1222

Table 2. Ephemerides of WDS 17190-3459 is shown in Table 3.

Table 1. Three fundamental points

| No | t | θ (º) | ρ (") | $\Delta_{qp}$ |
|---|---|---|---|---|
| 1 | 1952 | 625.529 | 2.054 | -1.9768 |
| 2 | 1972 | 510.866 | 1.059 | 1.8790 |
| 3 | 1992 | 273.073 | 2.097 | 0.5655 |

Table 2. Thiele-Innes Constants and its errors

| Thiele-Innes Constant | Value |
|---|---|
| A | 0".231 ± 0".010 |
| B | 1".236 ± 0".010 |
| F | 1".474 ± 0".249 |
| G | – 0".728 ± 0".246 |

Table 3. Ephemerides of WDS 17190-3459

| No | t | ρ (") | θ (º) |
|---|---|---|---|
| 1 | 2007.000 | 1.505 | 209.389 |
| 2 | 2008.000 | 1.460 | 203.107 |
| 3 | 2009.000 | 1.415 | 196.426 |
| 4 | 2010.000 | 1.369 | 189.299 |
| 5 | 2017.237 | 0.530 | 79.424 |

## III. Physical Parameters

Here we present physical parameters of WDS 17190-3459. We use semimajor axis a and periode P that yielded in orbital parameters' determination. Visual magnitude and spectrum (which used to get bolometric corection of each star) of this system are shown on Table 4. By using these values and mass-luminosity relation we get physical parameters of WDS 17190-3459, as shown on Table 5.

Table 4. Visual magnitude and bolometric correction of WDS 17190-3459 (Sources: Hartkopf, 2001 and Cox, 1976)

| Star | V | Spectrum | BC |
|---|---|---|---|
| 1 | 6.37 | K3V | -0.35 |
| 2 | 7.38 | K5V | -0.72 |

## IV. Conclusion

By using Thiele-Innes method, orbital parameters of WDS 17190-3459 can be obtained. As shown in Figure 3 our result can be used to describe an orbit of WDS 17190-3459. We have also analysed physical parameters of WDS 17190-3459. As shown in Table 5 our result are consistent with other work (Söderhjelm, 1999) which use Gauss method.

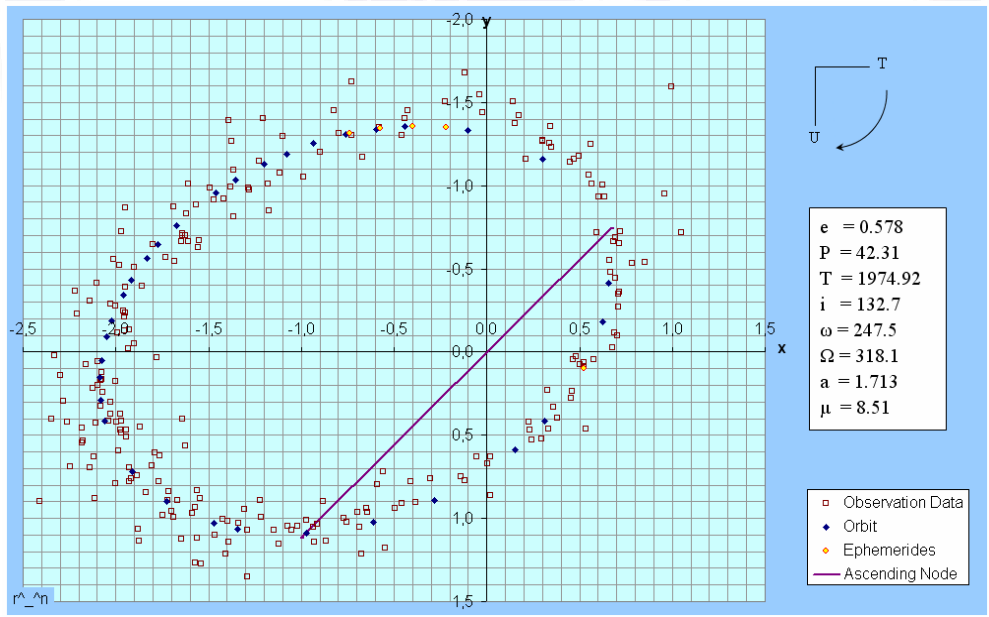

Figure 3. Orbit of WDS 17190-3459

Table 5. Comparison of author's (Thiele-Innes method) and Söderhjelm's (1999) (Gauss method)

| No | Parameters | | Authors | Söderhjelm (1999) |
|---|---|---|---|---|
| | **ORBITAL PARAMETERS** | | | |
| 1 | *Dynamical Elements* | | | |
| | Eccentricity | $e$ | 0.578 | 0.58 |
| | Period | $P$ | 42.31 | 42.15 |
| | Passage periastron | $T$ | 1974.92 | 1975.9 |
| 2 | *Orientation Elements* | | | |
| | Inclination | $i$ | $132°.7$ | $128°$ |
| | Argument periastron | $\omega$ | $247°.5$ | $247°$ |
| | Ascending node | $\Omega$ | $318°.1$ | $313°$ |
| 3 | *Geometric Ellipse* | | | |
| | Semi major axis | $a$ | 1".713 | 1".81 |
| | Semi minor axis | $b$ | 1".398 | 1".47 |
| 4 | *Annual mean motion* | $\mu$ | $8°.52$ | $8°.54$ |
| | **PHYSICAL PARAMETERS** | | | |
| 1 | Paralax | $p$ | 0".134 | - |
| 2 | Abslt. bol. mag. *1 | $M_{bol1}$ | 6.66 | - |
| 3 | Abslt. bol. mag. *2 | $M_{bol2}$ | 7.36 | - |
| 4 | Total Masses | $M_1 + M_2$ | 1.16 | 1.27 |
| 5 | Mass ratio | $q$ | 0.863 | 0.88 |

## VI. Acknowledgements

Thank you to WDS team who gave us a compilation of observational data of WDS 17190-3459. RN is grateful to Susan Febriantina who helped him during the preparation of this work.